\documentclass[namedreferences]{kluwer}

\usepackage{graphicx}

\def\'#1{\ifx#1i{\accent"13\i}\else{\accent"13#1}\fi}

\def\apj{{ApJ}}
\def\apjs{{ApJS}}
\def\apjl{{ApJL}}

\def\aap{{AA}}
    \def\mnras{{MNRAS}}
\def\nat{{Nature}}
\def\araa{{ARAA}}
  \def\pasj{{PASJ}}
     
\def\alamenos#1{$^{-#1}$}

\def\be{\begin{equation}}
\def\ee{\end{equation}}

\def\kms{{km s$^{-1}$}}

\def\diezala#1{10$^{#1}$}

\def\prommath#1{\langle #1\rangle}

\begin{document}

\begin{opening}

\title{Molecular Clouds. Formation and Disruption}

\author{\surname{Javier Ballesteros-Paredes}}  

\institute{Instituto de Astronom\'ia, UNAM
\email{j.ballesteros@astrosmo.unam.mx}}

\runningauthor{J. Ballesteros-Paredes}

\runningtitle{Molecular Clouds. Formation and Disruption}

\begin{ao}\\
Javier Ballesteros-Paredes \\
Instituto de Astronom\'ia, UNAM \\
Ap. Postal 372, C.P. 58089 \\
Morelia Michoac\'an \\
M\'exico \\

\end{ao}

\begin{abstract}

Molecular clouds (MC) are the densest and coldest component of the
interstellar gas, and the sites of star formation.  They are also
turbulent and fractal and their masses and sizes span several orders
of magnitude. It is also generally believed that they are close to
Virial equilibrium (VE). Since this statement has been questioned by a
number of authors, with important implications on molecular clouds'
lifetimes, we will review this subject within the context of a
turbulent ISM. In this framework, there is significant numerical
evidence that MCs are not in VE, that there is a strong exchange of
mass, momentum and energy between clouds and their surrounding medium,
and that it is difficult (if not impossible) to form quasistatic cores
inside MCs, suggesting that they must be transient, short-lived
phenomena. Thus, their formation and disruption must be primarily
dynamical, and probably not due to just a single mechanism, but rather
to the combination of several processes. This picture seems consistent
with recent estimates of ages of stars in the solar neighborhood.  

\end{abstract}


\end{opening} 

\section{Virial Theorem: are clouds actually in Virial
Equilibrium?}\label{VT:sec} 

The Virial theorem can be derived from the momentum equation by
dotting it by the position vector and integrating it over the volume
of interest (see, e.g., \opencite{Parker79}; \opencite{Spitzer82};
\opencite{Shu89}; \opencite{Hartmann98}). Due to lack of space, we
will not write down the equation and their terms. A clear and full
description of either the Lagrangian and Eulerian versions  can be
found in \inlinecite{MZ92} (see also \opencite{BV97};
\opencite{BVS99}). Here we just mention that it involves the
calculation of all the energies of the clouds, as well as the
evaluation of pressures at the clouds' surface, and time derivatives
of the moment of inertia. In the Eulerian version, additional terms
involving mass flux through the clouds' surfaces appear. 

It is frequently encountered in the literature that VE is applicable
to MCs (e.g., \opencite{McKee00} and references therein).
This statement is generally based on the theoretical assumption that
all forces are in balance (\opencite{Spitzer82}), or on the
observational fact that clouds exhibit near equipartition between
gravity and internal energies. Nevertheless, in a constantly stirred 
turbulent ISM (see chapter by Mac Low, this volume), it is not
clear that the forces will balance. Furthermore, equipartition between
energies should not be considered evidence for VE. The only probe
that VE admits is to measure the second time derivative of the moment
of inertia, $\ddot{I}$, and check that it is zero. Such a probe is
not possible in an observational way. 
However, it has been measured in numerical simulations of modeled
clouds by \inlinecite{BV97}, who have
shown that, for an ensemble of clouds in  simulations of the
ISM, the clouds are {\it not} in VE. They obey
the Virial Theorem, but the second time derivative of the moment of
inertia is {\it not} zero.  

The VE assumption ($\ddot I=0$) implies either that clouds are not
redistributing their mass inside, or that the variation of their
mass distribution is constant with time. Both statements seem highly
implausible in a dynamical, nonlinear ISM. \inlinecite{McKee00}
has suggested two possibilities, in order to assume VE  for MC: One is
that, for a single cloud, $\prommath{\ddot I}=0$
if the considered averaging time is much larger than the dynamical
timescale of the cloud, $t_{\rm avg} \gg t_{\rm dyn}$. The other is
that for an ensemble of clouds, some of
them may have positive values of $\ddot I$, and some others will have
negative values, so that VE holds for the ensemble.
Regarding the first assumption, there is a hypothesis behind it: the
cloud has to be oscillating around a mean shape without a strong
redistribution of mass. Nevertheless, in a nonlinear regime, any
variable may change in an arbitrary way, and not necessarily the mean
values will average to zero. Furthermore, recent numerical and
observational evidence (cf. \S 
\ref{formation:sec}) suggests that the lifetimes of the clouds are not
significantly larger than their own dynamical times (i.e., they are
transient), and averaging over large timescales may be
meaningless. Concerning the second assumption, the analysis by
\inlinecite{BV97} shows that variations
of the moment of inertia span up to 7 orders of magnitude (in absolute
value) between the largest clouds and the smallest. Thus, it is not
clear that an average for such large scatter will be representative of
the actual dynamics of the clouds. Moreover, even if $I$ averages to
zero for the cloud ensemble, this does not alter the fact that clouds
are individually {\it out} of VE.

It is important to mention that, however, for clouds in numerical
simulations, there is approximately an
energy equipartition between gravitational, kinetic,  magnetic and
internal energies \cite{BV95}, in agreement with observational
determinations (e.g., \opencite{Myers_Goodman88}), suggesting that, in
particular ``Virial mass'' estimates (which would be better called ``energy
equipartition mass'') are of the same order of magnitude than the
actual mass, in spite of VE probably not being applicable.

It is also widely believed that the turbulent kinetic
energy provides support for clouds against their self-gravity (see,
e.g., \opencite{Chandra51}, 
\opencite{Bonazzola87}, \opencite{vsg95}). However, this is only part
of the turbulent motions, and only if they were 
confined to scales much smaller than the cloud sizes it could be the
only effect
\cite{Chandra_Fermi53, Leorat_etal90, Padoan95, BVS99,
KHM00}. In reality, interstellar turbulence is a multiscale
phenomenon, and there 
is observational and numerical evidence that turbulence at very
different scales is present in MC, with most of the energy residing in
the large-scale motions, as in most turbulent flows
\cite{Scalo87, Norman_Ferrara96, Falgarone98, BVS99, Avila_Vazquez01,
OM02, Brunt02}. Moreover, turbulence in MCs is highly compressible,
and compressible turbulent modes have the effect of producing density
enhancements at scales smaller than their own, thus promoting, rather
than preventing, gravitational collapse. This process is usually
referred to as ``turbulent fragmentation'' (e.g.,
\opencite{Padoan95}). Thus, turbulent modes at a given scale are most likely
to have a dual role, providing support toward larger scales while
simultaneously promoting collapse of smaller ones \cite{KHM00}. The
precise outcome of 
collapse or support must be the result of 
the total energy involved in the compressible modes of turbulence at
large scales against the energy involved in the smaller scale modes.

On the other hand, it is found that surface and volumetric terms in
the Virial theorem are comparable in order of magnitude
(\opencite{BV97}, see also \opencite{BVS99}), suggesting that clouds
are actually interchanging mass, momentum and energy with the
surrounding medium. This result is reasonable because the ISM is
highly dynamic, with strong energy injection by stellar outflows,
spiral density waves,
HII region expansions, etc. In fact, thermal and 
kinetic pressure confinement has been criticized by
\inlinecite{BVS99}, arguing that thermal pressure
constancy is relatively meaningless in a medium in which the ram
pressure is significantly larger than the thermal one, and that ram
pressure ``confinement'' is meaningless, because, when the motions are
at the scales of the cloud itself, the cloud is actually a (transient)
turbulent fluctuation, rather than a persisting entity that somehow
requires to be ``confined''. Recent
observations confirming this picture are presented by
\inlinecite{Jenkins_Tripp01} 

The preceding discussion questioning the assumption of VE for
molecular clouds has an analogue at the level
of their dense cores. Although in the standard models for star
formation, low-mass protostellar cores are assumed to be in
quasi-hydrostatic equilibrium (e.g., \opencite{Mouschovias91},
\opencite{SAL87}, and references therein), this view has begun to
change recently, especially among the proponents of a turbulent
scenario for molecular cloud structure and star formation (e.g.,
\opencite{Scalo87}; \opencite{Elmegreen93}; \opencite{Padoan95};
\opencite{VPP96}; \opencite{BVS99}; \opencite{KHM00};
\opencite{Padoan_Nordlund02}; \opencite{VBK02}; \opencite{VSB02}). For
instance, \inlinecite{BV99} have presented preliminary results
suggesting that the 
velocity structure of the  so-called ``coherent cores'' 
(dense molecular cores with a non-thermal velocity dispersion that
becomes constant with size at small -below few $\times$ 0.1 pc-
scales; \opencite{Goodman_etal98}), may be interpreted as the
signature of the collision between gas streams. 
Similarly, \inlinecite{Padoan_etal01}, have found a velocity
dispersion-column density relationship which appears in
shocked generated cores in their simulations, as well in
observed protostellar cores, suggesting
that protostellar cores must be formed by supersonic
turbulence. 

Furthermore, a particular argument against the possibility that cores
within MCs are in hydrostatic regime has been given by
\citeauthor{VSB02} (2002, see also \opencite{Tohline_etal87},
\opencite{Taylor_etal96}, \opencite{BVS99},
\opencite{VSB02}). Molecular clouds are nearly isothermal, and thus a
typical Bonnor-Ebert (BE) configuration 
\cite{Ebert55, Bonnor56}, in which a tenuous, hotter medium provides
confining pressure, cannot exist, because the core and its
surrounding medium are at roughly the same temperature. Thus, cores
must be extended, rather than truncated structures. Now, extended
equilibrium isothermal structures can be shown to always be
gravitationally unstable,\footnote{It is important to
mention that the preceding 
discussion is not in contradiction with the fact that stars are
objects in hydrostatic equilibrium within a turbulent medium, since
they do not correspond to quasi-isothermal flows. In stars, energy is
trapped since the opacity has increased, and the cooling time is about
\diezala{10} times the free-fall time. In order for a core formed by a
compression to reach equilibrium, the isothermalicity of the flow must
be released. Under MC conditions, this
requirement occurs until scales and densities of protostars are
reached \cite{BVS99}.}  and so, if a turbulent, dynamic compression
brings the core near a hydrostatic configuration, the core will simply
proceed to collapse if it reaches gravitational instability, or else
will re-expand back to the average mean density of the cloud. However,
cores that re-expand will be delayed by their own self-gravity,
spending more time in this process than if they proceed to collapse. This
result is consistent with the fact that MC 
typically contain more starless than star-forming cores 
(\opencite{Taylor_etal96, Lee_Myers99}; see also \opencite{Evans_99} and
references therein). 

Of course, the above discussion does not apply if the dense cores are
embedded in a hotter, more tenuous medium, which can then confine the
core. B68 is probably the best example of a core in this
situation  \cite{Alves_etal01,Hotzel_etal02a,Hotzel_etal02b}, although
there are some pieces of inconsistency: First, the cloud is not round,
or oval in a regular way. This makes implausible that it can be in
precise hydrostatic equilibrium.  Second, their recently reported
observations show motions of 0.25$-$0.5 the sound speed, suggesting
that the cloud is near but not precisely in hydrostatic equilibrium.
Third, the fitted Bonnor-Ebert profile by \inlinecite{Alves_etal01}
implies a temperature of 16~K, and, even at this temperature, marginal
instability. The actual temperature is a factor of
30-50\% lower ($\sim 10$~K, see \opencite{Hotzel_etal02a}), and thus
the thermal support is even lower. Then, the BE fit implies
not hydrostatic equilibrium, but rather, instability. Finally,
\inlinecite{BKV02} have found, over a sample of 120 cores on their 3
projections, that 50\% of the projections can be fitted succesfully by
a BE profile, in spite of not been in hydrostatic equilibrium. This
shows that fitting BE-like column density profiles to cores in
numerical simulations is not an unambiguous test for hydrostatic
equilibrium. 

\section{Cloud Formation and Destruction}\label{formation:sec}

The evidence reviewed in the last section that clouds 
have important fluxes of the physical quantities (mass, momentum, energy)
through their boundaries, and that they have important time
derivatives of their moment of inertia, suggests that they may be
much short-lived than 
previously thought, in line with suggestions by,
among others, \inlinecite{Sasao73}, \inlinecite{Hunter79},
\inlinecite{Hunter_etal86}, \inlinecite{Verschuur91},
\inlinecite{Elmegreen93}, \inlinecite{BVS99}, that clouds are
turbulent density fluctuations.

As has been discussed in previous reviews (e.g.,
\opencite{Elmegreen93}), it is most likely that there are  
more than a single mechanism for cloud formation. There is
more or less agreement that the earlier models of collisional
agglomeration of smaller ``cloudlets'' by, e.g., \inlinecite{Kwan79},
\inlinecite{Scoville_Hersh79},
\inlinecite{Cowie80}, do not reproduce the
observations \cite{Blitz_Shu80}. In particular, in those
models, the time required for building up a giant MC ($\sim 100$ Myr)
is so long that it is
difficult to either support a GMC against collapse, or avoid their
disruption by the internal HII regions and 
SN explosions. In addition to the fact that there is not
enough molecular material in the ``chaff'' to allow the formation of
GMCs via agglomeration \cite{Blitz_Williams00}, the picture of
``cloudlets'' flying ballistically might be inadequate, since clouds
are not a collection of isolated entities of gas, but an
interconnected network that has only  been made to look like spheres
due to the limitations of the  observations \cite{Scalo90}. 

Within the context of a turbulent ISM, clouds may form by the
convergence of turbulent flows 
at large scales. These might be probably produced by different
instabilities (Parker, thermal, gravitational, or magnetorational
[\opencite{Sellwood_Balbus99}]), the
passage of spiral density waves, swept up shells from SN remnants or simply
the general action of global turbulence, where no single apparent
mechanism is invoked, but the result of streams at different
velocities that collide (see \opencite{Elmegreen90}, 
\opencite{Elmegreen93}; \opencite{Blitz_Williams00}  for reviews). An
interesting alternative considered by \inlinecite{Pringle_etal01} is
that a large fraction ($\sim 50$\%) of the interstellar medium may be in
the form of (non-self gravitating) molecular gas which is too cold to
be detected. In their picture, we can only observe the molecular gas that is
illuminated by sufficiently nearby heating sources. It is compressed in spiral
shocks and only there a substantial fraction of the gas becomes
self-gravitating enough to initiate star formation. It is only at this
point, they argue, that we are able to see it.

In any event, an important constraint that should be satisfied for any cloud
formation model are the observed ages of newborn stars in MCs,
since they are the most direct measurement of a time scale in
MCs. The other time scale is the dynamical or crossing time, $\tau \sim
l/\Delta v$. Although it is applicable only in a statistical sense,
there may be strong discrepancies when compared to the ages of
stars. For example, for Taurus, ($l\sim 20 $ pc, $\Delta v\sim
2$ \kms), the dynamical timescale is of the order of 10~Myr, while the
ages of the young stars are only $\sim 2$ Myr (see, e.g.,
\opencite{BHV99} and references therein). 

It is important to note that there is
some controversy on the determination of ages in star-formation
regions. For example, \citeauthor{Palla_Stahler00} (2000, 
\citeyear{Palla_Stahler02}) argue that nearby MC have
been forming stars in the last 10 Myr or more, with a recent burst
of star formation. However, 
\inlinecite{Hartmann02} argues that their conclusions are skewed by a
statistically small sample of stars with masses larger than 1$
M_\odot$, and by biases in their birthline age corrections. He notes
that the picture by Palla \& Stahler (a) requires the last 1-2 Myr
to be a special epoch for most MCs; (b) implies that most
MCs are forming stars at extremely low rates, if any; and that
(c) the apparently oldest stars are systematically higher in mass,
implying that for most of a typical MC's star-forming epoch, the
Initial Mass Function was strongly skewed. Item (a) seems implausible,
and items (b) and (c) are contradicted by observations (see
\opencite{Hartmann02} and references therein).  A more plausible
explanation of the observations is that the ``tail'' of older stars is
really the result of including older foreground stars, as well as
problems with the isochrones calibration in the higher mass stars.  

\inlinecite{Blitz_Williams00} argue that giant MC must
live for some times 10 Myr based on the ages of stars in Orion quoted 
by \inlinecite{Blaaw64}. Nevertheless, more recent estimates
of the stellar ages \cite{Brown_etal94,Briceno_etal01} show that in
the Orion OB1a association, which has $\sim$ 10~Myr old stars, there is
virtually no molecular gas associated. Instead, in Orion 
OB1b, which contains large amounts of gas, the stars are about 1~Myr
old, a value close to the estimated for young stars in Taurus. 

\inlinecite{HBB01} tabulate the ages of stars in 13 nearby star-forming 
regions. For those regions with stars older than $\sim 5$ Myr, 
there is no molecular gas associated, suggesting that the time scales
for both cloud-  and star-formation are shorter even than the
dynamical time scale proposed by \inlinecite{Elmegreen00}.

The fact that several molecular regions exhibit synchronized star
formation with ages of the newborn stars much smaller than the
dynamical time suggests that some kind of external triggering must be
involved. \inlinecite{VPP95}, \inlinecite{BHV99} and
\inlinecite{HBB01} suggest that global turbulence may play a crucial
role. In their picture, large-scale flows powered by global events of
star formation 
along the Galaxy may collide, form clouds and stars rapidly, and then
dissipate. The idea is not new \cite{Blaaw64, Elmegreen_Lada77,
McCray_Kafatos87}, but although mechanisms like nearby HII regions in
expansion, or SN events, are not discarded, 
all these processes feed the global turbulence, and thus no
single mechanism need be directly responsible for the formation of any
particular cloud complex.

How fast can MCs be formed by the general turbulence?
\inlinecite{BHV99} show that clouds can be produced rapidly (in few Myr) by
the convergence of large-scale flows, evolving to high densities over
scales of tens of parsecs nearly simultaneously. This is because the
velocities involved are of the order of the velocity dispersion at the
large scales (several km s$^{-1}$), rather than of the internal
velocity dispersion of the clouds. Nevertheless, the 
exact cloud build-up time
may depend on how much mass the streams are carrying, how strong the 
compression is, the rate of cooling of the compressed (shocked)
region, the geometry of the compression, etc.  Even with the typical
ISM flow velocities $\sim 10$ \kms, it can take tens of Myr to
accumulate enough mass from the diffuse interstellar medium ($n\sim
1$cm\alamenos 3) to form a MC complex. However, a
necessary (though not sufficient) condition for the existence of
molecular material in the solar neighborhood is that it have
sufficient column density to effectively shield H$_2$ and CO from the
dissociating ultraviolet radiation of the diffuse Galactic field.
This requires a minimum column density in hydrogen atoms of roughly
\diezala{21} cm\alamenos 2 (see \opencite{Franco93} and references
therein). Thus, even if the process of building up material from
diffuse H~I takes a long time, the ``life'' time of the MC
in the solar neighborhood only begins once this
minimum column density is attained \cite{HBB01}. 

An important point to note is that the column density needed to allow
molecular gas to be formed is similar to that required for the MC to become
self-gravitating, under solar-neighborhood conditions ($\lesssim 1
A_{\rm v}$).  This may be the main reason why star formation is presently
occurring in virtually all MC complexes of significant
size within a kiloparsec from Sun \cite{HBB01}, but the situation for the
inner or outer galaxy may be quite different \cite{Cox_Franco86}.

Concerning the destruction of clouds, studies of other star-forming
regions in addition to Taurus (\opencite{Hartmann02}, and references
therein) in the solar neighborhood such as Cha I and IC 348 
\cite{Lawson_etal96, Herbig98} also provide relatively little evidence 
for large populations of stars older than few Myr, especially when
observational biases are eliminated. This
suggests that the molecular gas may be dispersed also in a few Myr, a
time scale consistent with the cluster survey results of
\cite{Leisawitz_etal89}. 

Several mechanisms for cloud-disruption have been proposed: SN
explosions, the action of massive stars (UV ionizing radiation and/or
winds), the lowering of shielding by an expansion of the cloud, the
action of internal and/or external turbulence, etc. As pointed 
out by \inlinecite{Franco_etal94}, while  SN explosions are
the products of late stages of evolution, and take some time (6 Myr or
more) to be ``turned on'', massive stars on their main sequence  may
power the clouds almost immediately. Thus, even if SN explosions were
the main mechanism powering global turbulence in the ISM (see Mac Low,
this volume), it is likely that the main agent disrupting clouds is
the massive OB stars. This view has been recently supported by
\inlinecite{Matzner02}, who has argued that the most efficient
mechanism for cloud destruction is photoionization in H~II regions,
even considering the combined effects of winds (from proto-,
main-sequence and evolved stars) and SNe. Once HII regions are
created, their expansion is responsible for 
ionizing and photodissociating all the environmental gas. 

At this point it is important to mention that a comparison between
observational data and numerical simulations with a star formation
prescription in which stars are formed in the  densest regions
\cite{VBR97} supports the view that clouds are dispersed by OB stars. 
Indeed, consider 
the case of Orion OB1 (Fig.~\ref{orion_sim}a), mentioned before.
OB1a region contains  $\sim$10 Myr-old stars but no
molecular gas. Instead, the OB1b association contains 1 Myr-old stars, and
large amounts of gas. A similar situation is found in the simulations
(Fig.~\ref{orion_sim}b), where the energy 
input is due to ``O stars'' that last 6 Myr. As can be seen, the
older, $5-8$  Myr stars\footnote{Note that stars older than 6 Myr are
fossils, and they are not contributing anymore to the dispersal of the
parent cloud.} (smaller crosses), are $10 - 20$ pc away from the gas. 

\begin{figure}[H]
\tabcapfont
\centerline{%
\begin{tabular}{c@{\hspace{0.5pc}}c}
\end{tabular}}
\caption{a.~~ Orion OB1 association. b.~~ Cloud in simulations from
\cite{PVP95}. Note that stars of more than 5 Myr old are 10 pc or more
far from the dense gas. The simulation was not intended to reproduce
the particular behavior of Orion.}
\label{orion_sim}
\end{figure}

Other mechanisms may also help in the dispersal of the clouds, and
contribute to the low star-formation efficiency, such as
ejections from massive stars \cite{Withworth79} and winds and outflows
(e.g., \opencite{Norman_Silk80}; \opencite{Leisawitz_etal89};
\opencite{Matzner_McKee00}). However, these are not considered as the
main mechanisms because they are considerably less energetic than the
HII regions and/or SN explosions. Recently, two other mechanisms have
been proposed (although not explored in detail): the lowering of
shielding \cite{HBB01}, and the same large-scale streams that creates
the clouds, may destroy them \cite{BHV99}. In the first case, if
clouds like Taurus (with a mean column density that corresponds to an
$A_{\rm V}\sim 1-2$, \opencite{Arce_Goodman99}) suffer an expansion of
surface area by a factor of 2 or 3, the column density will be reduced
by the same factor, allowing the destruction of molecular gas, with 
the consequent increase of temperature and, thus, of the Jeans length. In
the second case, although a small fraction of the mass in clouds formed by
turbulence may be considerably self-gravitating and susceptible for
collapse and star formation, most of the mass may be (marginally)
unbound and thus easily dispersed \cite{Shadmehri_etal01, VBK02}.
These mechanisms, as well as the first mentioned, require further study,
in order to quantify their efficiency in disrupting less massive
clouds as, e.g., Taurus, where no massive stars or HII regions are
available to contribute to the dispersal. In any event, rapid
dispersal of gas is required to avoid extended periods of star
formation, as stellar ages suggest. 

\section{Conclusions}

We have discussed the implications of the Virial theorem on MCs'
lifetimes. Consisting in two dimensions of spatial information, 
one dimension of velocity information, and no time evolution
information, observations can not probe that clouds are or not in
Virial equilibrium. Thus, numerical simulations can show important
qualitative features, even if details were missing or
incorrect. Numerical works have shown that clouds are not in Virial
equilibrium, that there is strong exchanges of mass, momentum and energy
with their environment, that time derivatives are non-negligible, and
that surface terms can not be 
neglected, being always comparable to the respective volumetric
term. Nevertheless, these simulations show equipartition between the
self-gravity and the internal energies of the clouds, just as
observations do (e.g.,  $GM/R \sim \delta v^2$).

If time derivatives and surface terms in the Virial theorem are
important, then clouds (and cores) must be transient. We discussed
that several mechanisms may be responsible for the formation of
clouds, with some emphasis on the global effects of multiscale
turbulence. This picture is consistent with recent estimations of ages
of stars associated to molecular gas shown in the
literature. Regarding destruction of MCs, it is favored the scheme
in which HII regions have enough power, and act before SN, to be the 
dominant factor of cloud dispersal. Nevertheless, some other
mechanisms may operate, in order to understand the lack of extended
periods of star formation in low-mass starforming regions as Taurus.

\begin{acknowledgements}

I want to thank M. de Avillez and D. Breitschwerdt for their
invitation. L. Hartmann and E. V\'azquez-Semadeni for careful reading
of the manuscript and useful comments. C. Brice\~no for allow me to
publish Fig.~1a before publication. I acknowledge financial support
by CONACYT's grant and I39318-E. This work has made extensive use of
NASA's Astrophysics Data System Abstract Service.

\end{acknowledgements}

\end{document}